\documentclass[12pt]{article}
\pdfoutput=1
\usepackage{amsmath,amssymb,color}
\usepackage[utf8]{inputenc}
\usepackage[T1]{fontenc}
\usepackage{graphicx}
\usepackage{amsmath}
\usepackage{amssymb}
\usepackage{tensor} 
\usepackage{braket}
\usepackage{comment}
\usepackage{mathtools}
\usepackage{caption}

\def\red#1{{\color{red} #1}}

\def\prg#1{\par\medskip\noindent{\bf #1}}  \def\ra{\rightarrow}
\newcounter{nbr}

       \def\ads3{{\rm AdS$_3$}}

\def\bull{\raise.25ex\hbox{\vrule height.8ex width.8ex}}
\def\Lie{{\cal L}\hspace{-.7em}\raise.25ex\hbox{--}\hspace{.2em}}

\def\ric{{Ric}}

\def\hook{\hbox{\vrule height0pt width4pt depth0.3pt
\vrule height7pt width0.3pt depth0.3pt
\vrule height0pt width2pt depth0pt}\hspace{0.8pt}}
\def\inn{\hook}

                \def\L{{\mit\Lambda}}

                   \def\n{\nu}
               \def\l{\lambda}
\def\vphi{\varphi}    \def\ve{\varepsilon}  

                 \def\om{\omega}
                  
\def\nab{\nabla}

\def\cO{{\cal O}}

\let\Pi\varPi

\def\cO{{\cal O}}

   \let\Pi\varPi

\def\nn{\nonumber}
\def\be{\begin{equation}}             \def\ee{\end{equation}}
\def\ba#1{\begin{array}{#1}}          \def\ea{\end{array}}
\def\bea{\begin{eqnarray} }           \def\eea{\end{eqnarray} }
\def\beann{\begin{eqnarray*} }        \def\eeann{\end{eqnarray*} }
\def\beal{\begin{eqalign}}            \def\eeal{\end{eqalign}}
\def\lab#1{\label{eq:#1}}             \def\eq#1{(\ref{eq:#1})}
\def\balign{\begin{align}}            \def\ealign{\end{align}}
\def\bsubeq{\begin{subequations}}     \def\esubeq{\end{subequations}}
\def\bitem{\begin{itemize}\vspace{-1pt} \setlength\itemsep{-4.5pt} }
  \def\eitem{\end{itemize}\vspace{-1pt} }

\def\aff#1{\vspace{-12pt}{\normalsize #1}}
\title{Near horizon symmetry of extremal spacelike-stretched black holes}
\author{ B. Cvetkovi\'c and D. Rakonjac\footnote{
        Email addresses: \texttt{cbranislav@ipb.ac.rs, danilo.rakonjac@ipb.ac.rs}} \\
\aff{Institute of Physics, University of Belgrade,
                           Pregrevica 118, 11080 Belgrade, Serbia} }
\date{}

\begin{document}

\maketitle

\begin{abstract}
We analyze the near horizon structure of the extremal spacelike stretched black holes,  exact solutions of topologically massive gravity. We show that the algebra of improved canonical generator is realized as a single centrally extended Virasoro algebra. We obtain the entropy of the solution by using the Cardy formula and compare the results with the corresponding non-extremal case.
\end{abstract}

\section{Introduction}
Topologically massive gravity (TMG) is an extension of general relativity with cosmological constant by adding the gravitational Chern-Simons term to the action \cite{tmg}. For the negative values of the cosmological constant, this theory posseses interesting solutions, namely the maximally symmetric AdS\textsubscript{3} solution, and the BTZ black hole \cite{btz}. While general relativity in three dimensions (3D) is a topological theory, TMG is a dynamical theory, it posseses a propagating degree of freedom, the massive graviton \cite{carlip}. However, these solutions are plagued with serious issues. For usual sign of the gravitational coupling constant, the massive excitations around the AdS\textsubscript{3} have negative energy, rendering such a ground state unstable. Changing the sign of $G$ gives BTZ black hole negative energy \cite{moussa-2003,li-2008}. In order to solve this issue, it was proposed that instead, a warped AdS\textsubscript{3} vaccuum should be used as a possible stable ground state of the theory \cite{anninos, guica}.

The warped AdS\textsubscript{3} is a solution of TMG in which the symmetry group $SL(2,R) \times SL(2,R)$ is reduced to $SL(2,R) \times U(1)$. Observing the AdS\textsubscript{3} as a fibration of AdS\textsubscript{2}, the warped solution is obtained by stretching or squashing along timelike or spacelike fibres \cite{bengtsson}. Eliminating the possibilities of closed timelike curves, which were found in case of timelike warping \cite{anninos}, we focus on the spacelike-stretched solutions. By using topological identifications, a spacelike-stretched black hole solution can be obtained from spacelike-stretched AdS\textsubscript{3}, in a similar manner to how BTZ black hole can be obtained from the regular AdS\textsubscript{3} spacetime. These black holes are the subject of our investigation.

Namely, Anninos et al. \cite{anninos} investigated the thermodynamic properties of these solutions, and posed a hypothesis that they could be dual to a two-dimensional conformal field theory on the boundary. The asymptotic symmetries of warped AdS\textsubscript{3} were investigated by Comp\'ere et al. in \cite{compere}. Using the canonical formalism as a natural way to investigate the asymptotic symmetries of a dynamical system, Blagojevi\'c and Cvetkovi\'c \cite{mb2009} confirmed the hypothesis made by Anninos et al. and obtained the gravitational entropy from the central charges of the canonical algebra of asymptotic symmetry generators. This method is rooted in the idea of defining the black hole entropy as a conserved charge on the horizon \cite{wald-1993,mbbc-2019}. However, this method breaks in the extremal case, since then the Hawking temperature tends to zero, and the first law of black hole mechanics is identically satisfied.

In this paper, we resolve this issue by calculating the black hole entropy of an extremal black hole by investigating its near horizon limit. In analogy to Kerr/CFT correspondence investigated in \cite{guica-2008} , we obtain the near horizon geometry of an extremal spacelike-stretched black hole, and investigate its asymptotic symmetry structure using the first order canonical formalism developed in \cite{rakonjac}. After introducing a consistent set of asymptotic boundary conditions, we obtain the asymptotic symmetry group in the form of Virasoro algebra, which is different from the non-extremal case where a product of Kac-Moody and Virasoro algebra is obtained \cite{mb2009}. Using the method developed in the seminal paper of Brown and Henneaux \cite{brown}, we then obtained the central charges of the canonical algebra, from which we find the entropy using the Cardy formula. The solution for the entropy is not continuously related to the result of the non-extremal case, however, this is a particular feature of the difference in asymptotic algebra between the near horizon geometry of the extremal solution, and the usual non-extremal solution. Thus, we have completed the investigation started in \cite{mbbc-2019}, confirming that the first order canonical formalism can be used to calculate the entropy even in the extremal case.

The paper is organized as follows. In section 2, we review the Lagrangian formulation of TMG, written in tetrad formulation of Poincar\'e gauge theory, the field equations and the basic variables of the theory. Further, we discuss the form of the space-like stretched black hole solution in the extremal case, and using the appropriate limiting procedure, we obtain the near horizon solution which we focus on in later sections. In section 3, we define the consistent asymptotic conditions for the near horizon metric, from which we obtain the corresponding asymptotic conditions for the triad and connection. Inspection of symmetries that preserve these asymptotic conditions gives a result that the asymptotic symmetry group is the 2d conformal group. This is different from the symmetry of the non-extremal case, but this difference is indeed expected
\cite{guica-2008,sen}.  In section 4, we obtain the canonical realization of the asymptotic symmetry group. Using the general formula for improving of the canonical generator, obtained in \cite{mbbc-2019}, we find the conserved charge and the central charge algebra of improved generators. The conserved and central charge are then used in Cardy's formula to obtain the result for the entropy of an extremal spacelike-stretched black hole. In section 5, this result is discussed and compared to already established results given in \cite{mb2009}.

The notational conventions used in the paper are: the Latin indices denote the components with respect to the local Lorentz frame, while Greek indices denote the components with respect to the coordinate frame. The local Lorentz metric is taken with the signature: $\eta_{ij} = \rm diag(+,-,-)$. Totally antisymmetric Levi-Civita symbol is normalized to $\varepsilon^{012}=1$.

\section{Extremal spacelike stretched black holes and their near horizon geometry}

\par Topologically massive gravity with cosmological constant can be naturally recast in the formalism of Poincar\'e gauge theory of gravitation \cite{mb-2002}. The fundamental dynamical variables are the triad $b^i$ and the spin connection $\omega^{ij} = -\omega^{ji}$ (1-forms), while  their corresponding field strengths are the torsion $T^i$ and curvature $R^{ij}$ (2-forms). In 3D spacetime, it is convenient to rewrite the connection and curvature in terms of their duals, which are given by the rule $A^{ij} := -\varepsilon^{ij}_{\hphantom{ij}k}A^k$ where $A^{ij}$ is any antisymmetric 2nd tensor(in Lorentz indices). Thus\red{,} we obtain the dual connection $\omega^i$ and curvature $R^i$. In terms of these variables, the field strengths are given by formulas $T^i = \nab b^i := db^i + \ve^i_{\hphantom{i}jk}\omega^jb^k$ and $R^i = d\omega^i +\frac{1}{2}\varepsilon^i_{\hphantom{i}jk}\omega^j\omega^k$. The wedge products between forms are omitted for brevity.

The underlying geometric structure of the theory corresponds to  Riemann-Cartan geometry, with triad fields relating to orthonormal coframe fields, $g = \eta_{ij}b^i \otimes b^j$ being the metric, and $\omega^i$ being (the dual of) Cartan connection, while $T^i$ and $R^i$ correspond to Cartan torsion and curvature, respectively. For $T^i = 0$, the geometry reduces to standard Riemannian geometry in 3 dimensions.

\subsection{Lagrangian of TMG and field equations}

The TMG Lagrangian is defined by:
\be
	L = 2ab^iR_i - \frac{\L}{3}\ve_{ijk}b^ib^jb^k + \frac{a}{\mu}L_{CS}(\om) + \l_iT^i
\ee
where $a = \frac{1}{16\pi G}$, $\L < 0$ is the cosmological constant, $\mu$ is the graviton mass, $L_{CS}(\omega) = \omega^id\omega_i + \frac{1}{3}\varepsilon_{ijk}\omega^i\omega^j\omega^k$ is the Chern-Simons Lagrangian and $\lambda_i$ is the Lagrange multiplier which ensures the validity of the torsion constraint $T^i = 0$.

By varying the action $S=\int L$ with respect to $b^i$, $\omega^i$ and $\lambda^i$, we obtain the field equations. {\it After} using the third equation $T^i = 0$, we can write the first two equations as:
\bsubeq
\bea
&&2aR_i -\Lambda\varepsilon_{ijk}b^jb^k + \frac{2a}{\mu}C_i = 0\,, \\
&&\lambda_i = \frac{2a}{\mu}L_i\,,
\eea
\esubeq
where the Schouten 1-form $L^i$ is given by:
\be
	L^i = (Ric)^i - \frac{1}{4}Rb^i\,.
\ee
Here, Ricci 1-form is given by $(Ric)^i = \varepsilon^{ijk}h_j {\inn} R_k$ while scalar curvature is $R = h_i{\inn}(Ric)^i$. The Cotton 2-form is defined by $C^i = \nabla L^i := dL^i + \varepsilon^i_{\hphantom{i}jk}\omega^jL^k$.

{TMG possesses an interesting solution  spacelike strethced black hole. Let us now briefy discuss the basic features of the afore mentioned solution in the extremal case.}

\subsection{Extremal spacelike stretched black holes}

Spacelike stretched black hole is a  solution of TMG, obtained as a discrete quotient of spacelike stretched  vacuum \cite{anninos}. The asymptotic behaviour and black hole thermodynamics were investigated in \cite{mb2009}. The existence of asymptotic conformal symmetry was also
discovered and studied in this sector of TMG.

After introducing a more convenient notation for the parameters:
\be
\L = -\frac{a}{\ell^2} \qquad \nu = \frac{\mu\ell}{3}
\ee
we construct the metric of spacelike stretched black hole following the procedure of \cite{anninos}. We find that in Schwarzschild-like coordinates $(t, r, \vphi)$ it takes the form:
\bsubeq
\be
ds^2 =N^2dt^2 - \frac{dr^2}{B^2} - K^2(d\varphi + N_\varphi dt)^2
\ee
where we have:
\bea
&&N^2 = \frac{(\nu^2 + 3)(r-r_+)(r-r_-)}{4K^2}\,, \qquad B^2 = \frac{4N^2K^2}{\ell^2} \\
&&K^2 = \frac{r}{4}[3(\nu^2 - 1)r + (\nu^2+3)(r_+ + r_-) - 4\nu\sqrt{r_+r_-(\nu^2 + 3)}]\,,\\
&&N_\vphi = \frac{2\nu r - \sqrt{r_+r_-(\nu^2 + 3)}}{2K^2}\,.
\eea
\esubeq
The solution exists in the sector $\nu^2>1$, while $r_\pm$ are the coordinates of the outer and inner horizon, respectively.\\
The extremal solution is defined by the condition $r_+ = r_-$, wherefrom we get
\bsubeq
\bea
&&N^2 = \frac{(\nu^2 + 3)(r-r_+)^2}{4K^2}\,, \qquad B^2 = \frac{4N^2K^2}{\ell^2}\,, \\
&&K^2 = \frac{r}{4}[3(\nu^2 - 1)r + 2(\nu^2+3)r_+ - 4\nu r_+\sqrt{\nu^2 + 3}]\,,\\
&&N_\varphi = \frac{2\nu r - r_+\sqrt{\nu^2 + 3}}{2K^2}\,.
\eea
\esubeq

The specific feature of the extremal solution is the possibility of construction of the near-horizon geometry. The construction
is achieved by perfoming the following transformation of coordinates:
\be
	t = \frac{\tilde{t}}{\varepsilon r_+}\frac{2K_+\ell}{\nu^2 + 3} \qquad r = r_+(1+\varepsilon\tilde{r}) \qquad \varphi = \tilde{\varphi} - \frac{2\ell}{\nu^2 + 3}\frac{\tilde{t}}{\varepsilon r_+}
\ee
and performing the subsequent limit $\ve \ra \infty$, where
\begin{equation*}
	K_+ := K(r_+) = \frac{r_+}{2}[2\nu - \sqrt{\nu^2 + 3}]
\end{equation*}
After applying the previous transformation, we find the metric of near-horizon extremal spacelike stretched black hole:
\be
	ds^2 = \frac{\ell^2}{\nu^2 + 3}\left(\tilde{r}^2d\tilde{t}^2 - \frac{d\tilde{r}^2}{\tilde{r}^2}\right) - \left(K_+ d\tilde{\varphi} - \frac{2\nu\ell}{(\nu^2 + 3)}\tilde{r}d\tilde{t}\right)^2
\ee
\prg{Triad fields, connection and curvature.} Since the metric is given in a diagonal form, the orthonormal coframe can be chosen straighforwardly
\begin{equation}
\begin{split}
	&b^0 = \frac{\ell}{\sqrt{\nu^2 + 3}}\tilde{r}d\tilde{t} \qquad b^1 = \frac{\ell}{\sqrt{\nu^2 + 3}}\frac{d\tilde{r}}{\tilde{r}} \\
	&b^2 = K_+ d\tilde{\varphi} - \frac{2\nu\ell}{(\nu^2 + 3)}\tilde{r}d\tilde{t}
\end{split}
\end{equation}
The Levi-Civita connection is obtained from Cartan's structure equation $db^i + \varepsilon^i_{\hphantom{i}jk}\omega^jb^k = 0$:
\be
\om^0 = \frac{\nu}{\ell}b^0 \qquad \omega^1 = \frac{\nu}{\ell}b^1 \qquad \omega^2 = -\frac{\sqrt{\nu^2 + 3}}{\ell}b^0 - \frac{\nu}{\ell}b^2
\ee
Finally using the definition of curvature, we find curvature 2-forms:
\be
R^0 = -\frac{\nu^2}{\ell^2}b^1b^2 \qquad R^1 = -\frac{\nu^2}{\ell^2}b^0b^2 \qquad R^2 = -\frac{2\nu^2-3}{\ell^2}b^0b^1\,.
\ee
 Then, the  Ricci 1-form $(Ric)^i = \varepsilon^{ijk}h_j \inn R_k$ is given by:
\bsubeq
\bea
&&(\ric)^{0}=\frac{3-\n^2}{\ell^2}\, ,\qquad(\ric)^{1}=\frac{3-\n^2}{\ell^2}\,,\qquad(\ric)^{2}=\frac{2\n^2}{\ell^2}\,.
\eea
\esubeq
Finally, the Cotton 2-form reads:
\bea
&&C_0=\frac{3\n}{\ell^3}(\n^2-1)b^1b^2\, ,\quad C_1=\frac{3\n}{\ell^3}(\n^2-1) b^2b^0\nn\,,\quad
C_2=-\frac{6\n}{\ell^3}(\n^2-1)b^0b^1\, .
\eea

As expected the field equations are exactly satisfied.

\section{Asymptotic conditions}
\setcounter{equation}{0}

After defining the near-horizon geometry of an extremal spacelike stretched black hole, we proceed to formulate the asymptotic boundary conditions in the near horizon region. The  transformation that we have performed to define the near horizon geometry is not a simple coordinate transformation, due to the limit which was takem at the end of the procedure. Because of this limit, the transformation is singular, and the resulting spacetime is not diffeomorphic to the spacetime we had started with. A similar situation was observed in \cite{rakonjac}. Since the geometry is not asymptotically flat, it is not obvious which boundary conditions are supposed to be imposed. In case of three-dimensional warped geometries, the asymptotic symmetries were investigated in \cite{compere}. In GR, this result has been obtained for different geometries \cite{guica-2008,compere-kerrcft}. Using these approaches as a reference, we will look to obtain the consistent asymptotic symmetries for the metric.
 Moreover, the asymptotic conditions of the triad fields are not precisely governed by the asymptotic conditions imposed on the metric, so a consistent choice of triad as well as near horizon conformal symmetry that appears will be shown in this section.
 
\prg{Metric asymptotics.} We introduce the following set of asymptotic boundary conditions for the metric at $\tilde{r}\rightarrow \infty$:
\begin{equation}
	\delta g_{\mu\nu} = \begin{pmatrix}
							\mathcal{O}_{-2} & \mathcal{O}_{2} & \mathcal{O}_{0} \\
							 \cO_2& \mathcal{O}_{4} & \mathcal{O}_{1} \\
							\cO_0 &\cO_1  & \mathcal{O}_{0}
						\end{pmatrix}
\end{equation}
Here we used the notation $\mathcal{O}_n := \mathcal{O}(\tilde{r}^{-n})$. It can be observed that these conditions slightly differ from the usual assumption of boundary conditions containing first subdominant terms to the metric. The result we have is closely analogous to the result obtained in the 4D case in \cite{guica-2008, rakonjac}.

\prg{Triad fields.} The corresponding asymptotic form for the triad fields is:
\begin{equation}
	b^i_{\hphantom{i}\mu} = \begin{pmatrix}
							\mathcal{O}_{-1} & \mathcal{O}_{3} & \mathcal{O}_{1} \\
							\mathcal{O}_{1} & \bar{b}^1_{\hphantom{1}\tilde{r}} + \mathcal{O}_{3} & \mathcal{O}_{0} \\
							\frac{\bar{b}^2_{\hphantom{2}\tilde{t}}}{f(\tilde{\varphi})} + \mathcal{O}_{0} & \mathcal{O}_{3} & \bar{b}^2_{\hphantom{2}\tilde{\varphi}}f(\tilde{\varphi}) + \mathcal{O}_{1}
						\end{pmatrix}
\end{equation}
where the background triad fields are given by:
\begin{equation}
	\bar{b}^i_{\hphantom{i}\mu} = \begin{pmatrix}
							\frac{\ell}{\sqrt{\nu^2 + 3}}\tilde{r} & 0 & 0 \\
							0 & \frac{\ell}{\sqrt{\nu^2 + 3}}\frac{1}{\tilde{r}} & 0 \\
							- \frac{2\nu\ell}{(\nu^2 + 3)}\tilde{r} & 0 & K_+
						\end{pmatrix}
\end{equation}
and $f(\tilde{\varphi}) = 1 + h(\tilde{\varphi})$ is an arbitrary function, with $h(\tilde{\varphi}) \ll 1$.

\prg{Asymptotic symmetry.} The asymptotic form of the metric is preserved by asymptotic Killing vector $\xi^\mu$ of the following general form:
\begin{equation}
	\xi = (T - C\tilde{t} + \mathcal{O}_3)\partial_{\tilde{t}} + (-\tilde{r}(\varepsilon'(\tilde{\varphi}) + C) + \mathcal{O}_1)\partial_{\tilde{r}} + (\varepsilon(\tilde{\varphi}) + \mathcal{O}_2)\partial_{\tilde{\varphi}}
\end{equation}
where $T$, $C$ are arbitrary constants, and $\varepsilon(\tilde{\varphi})$ is an arbitrary function of $\tilde{\varphi}$.
\par The subdominant terms in the expression correspond to the trivial diffemorphisms and can be disregarded. Moreover, the transformations with $\varepsilon = 0$ represent the residual gauge transformations which give trivial contribution to the central charge, and are therefore not of interest to us. We form asymptotic symmetry group as a quotient with respect to the residual transformations. What remains is an asymptotic Killing vector which generates the conformal group of the circle:
\begin{equation}\lab{3.17}
	\xi = -\tilde{r}\varepsilon'(\tilde{\varphi})\partial_{\tilde{r}} + \varepsilon(\tilde{\varphi})\partial_{\tilde{\varphi}}
\end{equation}
From the general algebra of PG, we find that the composition of asymptotic transformations is of the form:
\begin{equation}\lab{3.18}
\begin{split}
	&[\delta_0(\varepsilon_1),\delta_0(\varepsilon_2)] = \delta_0(\varepsilon_3),\\
	&\varepsilon_3 = \varepsilon_1\varepsilon'_2 - \varepsilon_2\varepsilon'_1
\end{split}
\end{equation}
where $'$ represents derivative with respect to $\tilde{\varphi}$.\\\\
Rewritten in terms of the Fourier modes, $\ell_n = \delta_0(\varepsilon = e^{in\tilde{\varphi}})$, the algebra takes Virasoro form:
\begin{equation}
	[\ell_n , \ell_m ] = i(m-n)\ell_{m+n}
\end{equation}
The transformation law of triad fields under PG transformations reads:
\begin{equation}
	\delta_0 b^i_{\hphantom{i}\mu} = -\varepsilon^i_{\hphantom{i}jk}b^j_{\hphantom{j}\mu}\theta^k - (\partial_\mu \xi^\rho)b^i_{\hphantom{i}_\rho} -\xi^\rho \partial_\rho b^i_{\hphantom{i}\mu}
\end{equation}
where $\xi^\mu$ and $\theta^\mu$ are parameters of local translations and local Lorentz rotations. respectively. From the condition that the asymptotic form of the triad is preserved under these transformations, we find Lorentz parameters:
\begin{equation}
	\theta^0 = \mathcal{O}_2 \qquad \theta^1 = \mathcal{O}_1 \qquad \theta^2 = \mathcal{O}_2
\end{equation}
The spin connection is Riemannian, so it can be expressed in terms of the triads, and therefore its asymptotic form is preserved under transformations \eq{3.17}.

In what follows, we shall use the canonical approach to investigate the asymptotic symmetry, and calculate the central charge and entropy of the system.

\section{Conserved charge, central charge and entropy}
\setcounter{equation}{0}

In calculating the central charge and conserved charge on the horizon, we shall make use of the general formula for variation of the canonical generator on the horizon, developed in \cite{mbbc-2019}:
\bea\lab{4.1}
&&\delta \Gamma = \oint_{S_H}\delta B(\xi)\,,\nn\\
&&\delta B(\xi) := (\xi \inn b^i)\delta\tau_i + \delta b^i(\xi \inn \tau_i) + (\xi \lrcorner \omega^i)\delta\rho_i + \delta\omega^i(\xi \inn \rho_i)\,,
\eea
where $\rho_i = \frac{\partial L}{\partial R^i}$ and $\tau_i = \frac{\partial L}{\partial T^i}$ are the covariant momenta. The covariant momenta are easily obtained from the Lagrangian:
\begin{equation}
	\tau_i = \lambda_i = \frac{2a}{\mu}L_i \qquad \rho_i = 2ab_i + \frac{a}{\mu}\omega_i
\end{equation}
The explicit form of covariant momenta is given by:
\begin{equation}
\begin{split}
	&\tau_0 = \frac{a(-2\nu^2+3)}{3\nu\ell}b^0 \qquad \rho_0 = \frac{7a}{3}b^0 \\
	&\tau_1 = \frac{a(2\nu^2-3)}{3\nu\ell}b^1 \qquad \rho_1 = -\frac{7a}{3}b^1\\
	&\tau_2 = -\frac{a(4\nu^2-3)}{3\nu\ell}b^2 \qquad \rho_2 = \frac{a\sqrt{\nu^2+3}}{3\nu}b^0 -\frac{5a}{3}b^2
\end{split}
\end{equation}
\subsection{Conserved charge}
In this subsection we compute the conserved charge on the horizon. It is obtained by using the exact Killing vector $\xi = \partial_{\tilde{\varphi}}$ in the general formula \cite{mbbc-2019}.  The non-zero terms in the variation of the surface term read:
\begin{align}
	&b^2_{\hphantom{2}\tilde{\varphi}}\delta\tau_2 = \delta b^2\tau_{2\tilde{\varphi}} = -\frac{a(4\nu^2-3)(2\nu - \sqrt{\nu^2 +3})^2}{24\nu\ell}\delta[r_+^2]d\tilde{\varphi} \\
	&\omega^2_{\hphantom{2}\tilde{\varphi}}\delta\rho_2 = \delta\omega^2\rho_{2\tilde{\varphi}} = \frac{5a\nu(2\nu-\sqrt{\nu^2 +3})^2}{24\ell}\delta[r_+^2]d\tilde{\varphi}
\end{align}
From the expression above, we find that the conserved charge is:
\begin{equation}
	J = \oint_{S_H} b^2_{\hphantom{2}\tilde{\varphi}}\delta\tau_2 + \delta b^2\tau_{2\tilde{\varphi}} + \omega^2_{\hphantom{2}\tilde{\varphi}}\delta\rho_2 + \delta\omega^2\rho_{2\tilde{\varphi}} = \frac{a\pi(\nu^2+3)(2\nu-\sqrt{\nu^2+3})^2}{6\nu\ell}r_+^2
\end{equation}
\subsection{Central charge}
The central charge is obtained from the algebra of improved canonical generators which has the following form:
\begin{equation}
	\{\tilde{G}(\varepsilon_1),\tilde{G}(\varepsilon_2)\} = \tilde{G}(\varepsilon_3) + C
\end{equation}
where $\varepsilon_3$ is defined by the composition rule \lab{3.18}, and $C$ is the central extension of the algebra.

We shall make use of the general result of Brown and Henneaux \cite{brown} in order to simplify the algebra of canonical generators to the following sequence of weak equalities:
\begin{equation}
	\{\tilde{G}(\varepsilon_1),\tilde{G}(\varepsilon_2)\} \approx \delta(\varepsilon_1)\Gamma(\varepsilon_2) \approx \Gamma(\varepsilon_3) + C
\end{equation}
Since the central charge is a constant functional, it can be obtained by performing variations on the background configuration. The application of the general formula  gives:
\begin{equation}
	\Gamma = aK^2_+\frac{\nu^2+3}{3\nu\ell}\int_0^{2\pi}(\varepsilon_1\varepsilon'_2 - \varepsilon_2\varepsilon'_1)d\tilde{\varphi} -\frac{a\ell}{3\nu}\frac{5\nu^2 + 3}{\nu^2 + 3}\int_0^{2\pi}(\varepsilon'_1\varepsilon''_2 - \varepsilon'_2\varepsilon''_1)d\tilde{\varphi}
\end{equation}
We can identify the second term as the central charge, while the first term represents the surface term with parameter $\varepsilon_3 = \varepsilon_1\varepsilon'_2 - \varepsilon_2\varepsilon'_1$. For computational details see the Appendix A.\\
Thus we have obtained the central charge in the form:
\begin{equation}
	C = -\frac{a\ell}{3\nu}\frac{5\nu^2 + 3}{\nu^2 + 3}\int_0^{2\pi}(\varepsilon'_1\varepsilon''_2 - \varepsilon'_2\varepsilon''_1)d\tilde{\varphi}
\end{equation}
In terms of Fourier modes, the canonical algebra takes the form:
\begin{equation}
	\{L_n, L_m\} = -i(n-m)L_{m+n} - \frac{c}{12}in^3\delta_{n, -m}
\end{equation}
In string theory normalization, we have:
\begin{equation}
	c = 12 \cdot \frac{4a\ell\pi}{3\nu}\frac{5\nu^2 + 3}{\nu^2 + 3}
\end{equation}
Now the entropy is obtained using the Cardy formula:
\begin{equation}
	S = 2\pi\sqrt{\frac{cL_0}{6}} = \frac{4a\pi^2\sqrt{5\nu^2+3}}{3\nu}(2\nu-\sqrt{\nu^2+3})r_+
\end{equation}
\section{Discussion}

As we can observe, the result obtained from the near-horizon geometry differs from the extremal limit of the entropy obtained in \cite{mb2009} by a constant multiplicative factor. This discrepancy can be explained as follows. In \cite{mb2009}, the canonical realization of the asymptotic symmetry gives us a canonical algebra of improved generators that is a semi-direct product of Kac-Moody algebra and Virasoro algebra. Via Sugawara \cite{sugawara} construction, the Virasoro algebra is obtained from the Kac-Moody factor, giving a direct product of two Virasoro algebras, corresponding to the left- and right-moving sectors in the dual CFT. The near-horizon limit of the extremal black hole, meanwhile, gives only one Virasoro algebra as the canonical realization of asymptotic symmetries. This is something that is expected in near-horizon geometries \cite{sen}. Moreover, following through the Sugawara procedure in \cite{mb2009}, we see that the algebra that we have obtained corresponds directly to the right-moving sector of the non-extremal solution, while the extremal limit of the entropy obtained for the non-extremal solution shows a contribution only from the left-moving sector, which came from the Kac-Moody algebra. This difference in asymptotic symmetry manifests itself in the resulting entropy in the extremal case being different from taking the extremal limit of the non-extremal entropy that was obtained before.

\appendix
\section{Calculation of the central charge}
\setcounter{equation}{0}
As we mentioned in section 4, the central charge is obtained by calculating the variation of the boundary term $\delta(\varepsilon_1)\Gamma(\varepsilon_2)$ on the background configuration, using the asymptotic Killing vector given in (3.5).\\
We use the following non-vanishing interior products with triad and connection:
\begin{align}
	\xi \lrcorner \bar{b}^1 = - \frac{\ell}{\sqrt{\nu^2 + 3}}\varepsilon'(\tilde{\varphi}) \qquad \xi \lrcorner \bar{b}^2 = K_+\varepsilon(\tilde{\varphi})\\
	\xi \lrcorner \bar{\omega}^1 = -\frac{\nu}{\sqrt{\nu^2 + 3}}\varepsilon'(\tilde{\varphi}) \qquad \xi \lrcorner \bar{\omega}^2 = - \frac{\nu}{\ell}K_+\varepsilon(\tilde{\varphi})
\end{align}
Using these interior products, we derive the non-vanishing interior products with the covariant momenta:
\begin{align}
	&\xi \lrcorner \rho ^1 = -\frac{7a\ell}{3\sqrt{\nu^2+3}}\varepsilon'(\tilde{\varphi}) \\
	&\xi \lrcorner \rho^2 = \frac{5a}{3}K_+\varepsilon(\tilde{\varphi}) \\
	&\xi \lrcorner \tau^1 = -\frac{a(-2\nu^2+3)}{3\nu\sqrt{\nu^2+3}}\varepsilon'(\tilde{\varphi}) \\
	&\xi \lrcorner \tau^2 = \frac{a(4\nu^2-3)}{3\nu\ell}K_+\varepsilon(\tilde{\varphi})
\end{align}
Nonvanishing components of the variation of the background triad field, defined on the boundary $\tilde{t} = const.$ , $\tilde{r} \rightarrow \infty$ are given by:
\begin{equation}
	\delta_0 \bar{b}^1 = \varepsilon''(\tilde{\varphi})\frac{\ell}{\sqrt{\nu^2+3}}d\tilde{\varphi} \qquad \delta_0 \bar{b}^2 = -\varepsilon'(\tilde{\varphi})K_+ d\tilde{\varphi}
\end{equation}
Using these results we can calculate the variation of covariant momenta:
\begin{align}
	&\delta_0 \tau^1 = \frac{a(-2\nu^2+3)}{3\nu\sqrt{\nu^2+3}}\varepsilon''(\tilde{\varphi})d\tilde{\varphi} \qquad \delta_0 \tau^2 = -\frac{a}{3\nu\ell}(4\nu^2 - 3)K_+\varepsilon'(\tilde{\varphi})d\tilde{\varphi} \\
	& \\
	&\delta_0 \rho^1 = \frac{7a\ell}{3\sqrt{\nu^2+3}}\varepsilon''(\tilde{\varphi})d\tilde{\varphi} \qquad \delta_0 \rho^2 = -\frac{5}{3}aK_+\varepsilon'(\tilde{\varphi})d\tilde{\varphi}
\end{align}
Finally, the variation of the connection on the background configuration is given by:
\begin{equation}
	\delta_0 \omega^1 = \frac{\nu}{\sqrt{\nu^2+3}}\varepsilon''(\tilde{\varphi})d\tilde{\varphi} \qquad \delta_0 \omega^2 = \frac{\nu}{\ell}\varepsilon'(\tilde{\varphi})K_+ d\tilde{\varphi}
\end{equation}
Summing all these contributions and integrating them according to the formula (4.1), we find the boundary term that is given in section 4:
\begin{equation}
	\Gamma = aK^2_+\frac{\nu^2+3}{3\nu\ell}\int_0^{2\pi}(\varepsilon_1\varepsilon'_2 - \varepsilon_2\varepsilon'_1)d\tilde{\varphi} -\frac{a\ell}{3\nu}\frac{5\nu^2 + 3}{\nu^2 + 3}\int_0^{2\pi}(\varepsilon'_1\varepsilon''_2 - \varepsilon'_2\varepsilon''_1)d\tilde{\varphi}
\end{equation}

\section*{Acknowledgements}

This work was partially supported by the Ministry of Education, Science and Technological development of the Republic of Serbia.

\end{document}